# The surface accessibility of α-bungarotoxin monitored by a novel paramagnetic probe

Andrea Bernini[a,b], Vincenzo Venditti[a], Ottavia Spiga[a,b], Filippo Prischi[a], Mauro Botta[c], Angela Pui-Ling Tong[d], Wing-Tak Wong[d] and Neri Niccolai[a,b,*]

(a): Dipartimento di Biotecnologie, Università di Siena; (b): SienaBioGrafiX Srl, I-53100 Siena, Italy; (c): Dipartimento di Scienze dell'Ambiente e della Vita, Università del Piemonte Orientale, I-15100 Alessandria, Italy; (d): Department of Chemistry, The University of Hong Kong, Pokfulam Road, HKSAR, China.

* Corresponding author: prof. Neri Niccolai, Università degli Studi di Siena, Dipartimento di Biotecnologie, via A. Fiorentina 1, 53100 Siena, Italy; phone +39577234910, fax +39577234903, e-mail niccolai@unisi.it.

**Summary**

The surface accessibility of α-bungarotoxin has been investigated by using $Gd_2L7$, a newly designed paramagnetic NMR probe. Signal attenuations induced by $Gd_2L7$ on α-bungarotoxin CαH peaks of $^1H$-$^{13}C$ HSQC spectra have been analyzed and compared with the ones previously obtained in the presence of GdDTPA-BMA. In spite of the different molecular size and shape, for the two probes a common pathway of approach to the α-bungarotoxin surface can be observed with an equally enhanced access of both GdDTPA-BMA and $Gd_2L7$ towards the protein surface side where the binding site is located. Molecular dynamics simulations suggest that protein backbone flexibility and surface hydration contribute to the observed preferential approach of both gadolinium complexes specifically to the part of the α-bungarotoxin surface which is involved in the interaction with its physiological target, the nicotinic acetylcholine receptor.

**Introduction**

Nowadays, the use of engineered proteins as nanomachines is already included in projects for developing new materials and devices. However, programming the activity of proteins requires a deeper insight on the mechanisms of their intermolecular approaches. Mapping protein surface accessibility might give a solid experimental basis to predict surface hot spots and, hence, to design suitable mutants. Protein surface accessibility, indeed, controls all the biological processes which are driven by protein-protein, protein-ligand and protein-nucleic acid interactions. The fact that protein structures and functions have evolved in an aqueous environment suggests that protein surface accessibility is modulated by a variety of mechanisms where water molecules play a predominant role. In facts, water molecules have been found to modulate the approach of organic solvent molecules to specific protein surface regions both in the crystal [1] and in solution [2]. Thus, a direct investigation on the dynamic state of water molecules located at the solvent-protein interface would yield precious information on protein surface regions exhibiting enhanced accessibility, therefore, most likely including the so called surface hot spots. However, the fact that the water-protein interaction network cannot be unambiguously defined [2] suggests that alternative experimental strategies are required for delineating the surface accessibility of proteins. In this respect, NMR studies on the effects induced by soluble paramagnetic probes, such as aminoxyl spin-labels [3-5], gadolinium complexes [6, 7] and molecular oxygen [8, 9], on protein and RNA NMR signals have provided a good wealth of information about the complex dynamics contributing to surface accessibility. The fact that paramagnetic probes might be involved in biased approaches towards specific amino acid side chains or structural determinants has been considered and, consequently, the use of more than one probe has been suggested to enhance the resolution of this kind of studies [10]. Some interference to a purely random approach of TEMPOL to the protein surface, due to some hydrogen bonding between the N-oxyl moiety of the probe and protein backbone amide groups, has been suggested by comparing the paramagnetic profiles induced by different probes [10]. Relaxometric techniques have yielded no evidence of strong interactions of GdDTPA-BMA with specific molecular sites, even in the crowded molecular environments typical of biological fluids [11], suggesting that the latter paramagnet is well suited for accurate mapping of surface accessibility [6].

In the present study a newly designed neutral complex of Gd(III) ion, whose structure is shown in Fig. 1, and henceforth called $Gd_2L7$, has been used for investigating the role of size and hydrophobicity of paramagnetic probes in defining protein surface accessibility.

In this respect, a comparison has been made between the paramagnetic attenuations induced by the presence of this ditopic probe and of GdDTPA-BMA on $^1H$-$^{13}C$ HSQC well resolved CαH signals of α-bungarotoxin, α-BTX. This small neurotoxin, whose structure and dynamics have been well characterized [12, 13], represent a suitable model system since specific pathways of approach to the protein surface have been recently found by using GdDTPA-BMA as a paramagnetic probe, as well as a remarkable complementarity between predicted low water density and high paramagnetic attenuations [14]. The surface dynamics picture inferred by the latter NMR study results in very good agreement with the recent crystallographic structure of α-BTX bound with its receptor [15]. The protein moiety most accessible to GdDTPA-BMA is, indeed, also the one which is in contact with the α1 subunit of the mouse nicotinic acetylcholine receptor, AchR. Moreover, in the latter protein-protein interface no water molecules are present, consistently with the lack of high water density sites in the α-BTX binding region predicted by MD simulations [14].

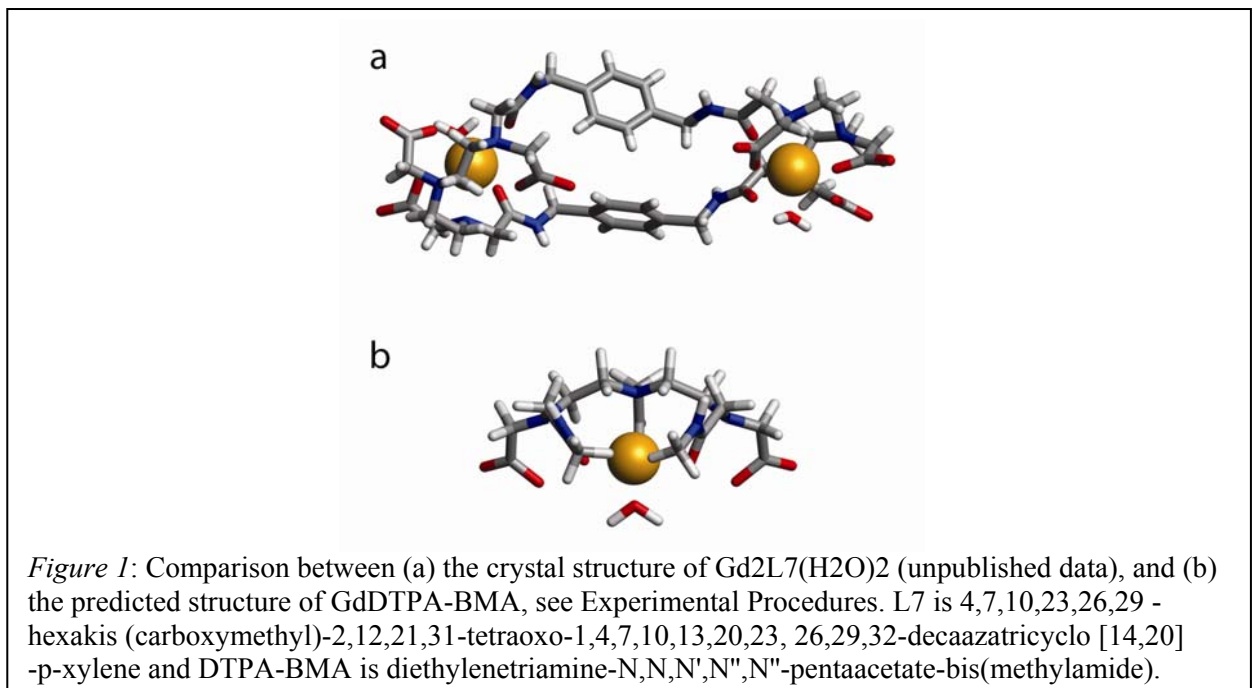

*Figure 1*: Comparison between (a) the crystal structure of Gd2L7(H2O)2 (unpublished data), and (b) the predicted structure of GdDTPA-BMA, see Experimental Procedures. L7 is 4,7,10,23,26,29 - hexakis (carboxymethyl)-2,12,21,31-tetraoxo-1,4,7,10,13,20,23, 26,29,32-decaazatricyclo [14,20] -p-xylene and DTPA-BMA is diethylenetriamine-N,N,N',N'',N''-pentaacetate-bis(methylamide).

In the case $Gd_2L7$ would confirm the accessibility pattern to the α-BTX surface already derived from GdDTPA-BMA paramagnetic perturbations, would suggest that the new ditopic probe should be included in the list of molecular tools for understanding protein surface dynamics.

**Results**

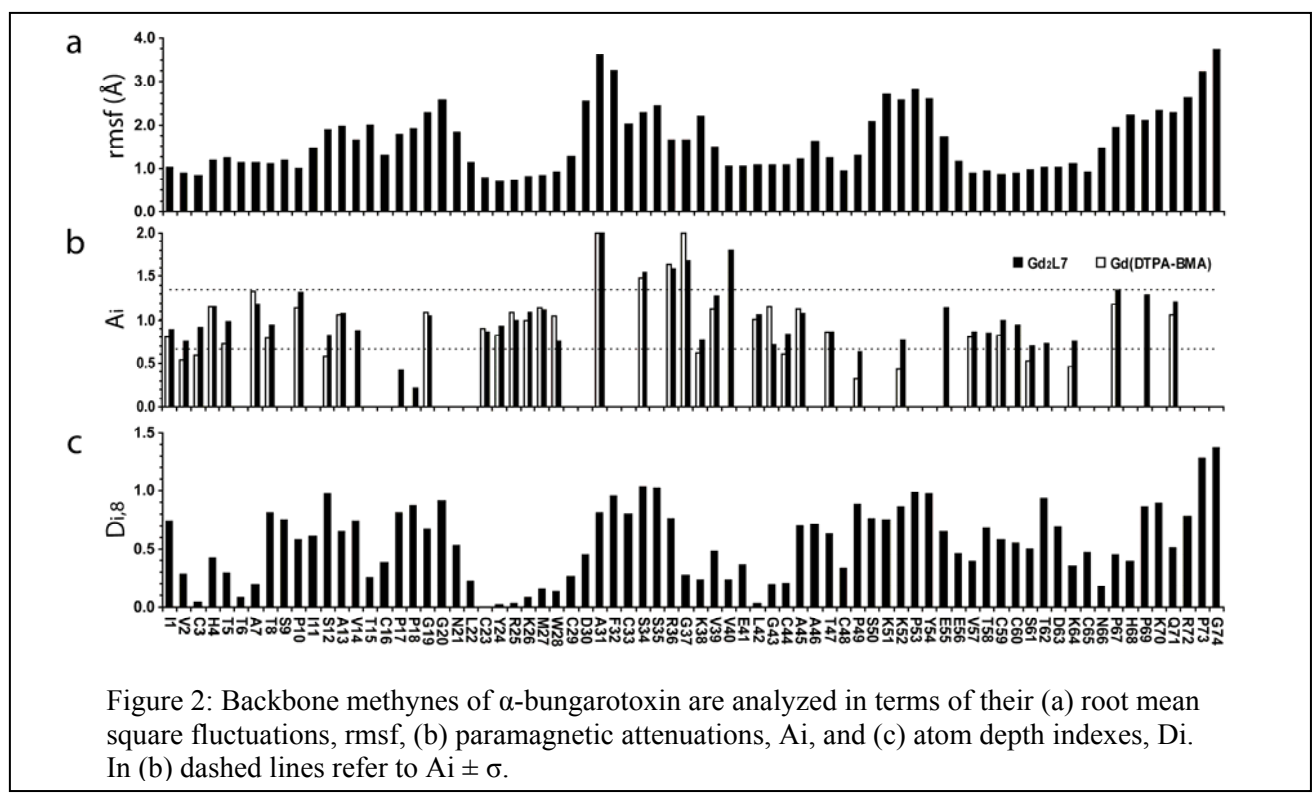

Figure 2: Backbone methynes of α-bungarotoxin are analyzed in terms of their (a) root mean square fluctuations, rmsf, (b) paramagnetic attenuations, $A_i$, and (c) atom depth indexes, $D_i$. In (b) dashed lines refer to $A_i \pm \sigma$.

*Molecular Dynamics simulation of α-BTX.* In order to correlate local molecular flexibility, surface hydration and protein surface accessibility, a MD simulation has been performed on α-BTX. As shown in Fig. 2a, residues 30-37 belonging to the so called α-BTX finger II and to the protein carboxy terminus, residues 67-74, are very flexible protein moieties, as inferred by their high root mean square fluctuation (rmsf) values. It should be noted that both regions are primarily involved in the binding with the AchR [15], and that in the latter side of α-BTX no MD defined high water density sites could be found [14]. This finding supports the already suggested correlation between the extent of protein backbone flexibility and absence of structured water molecules [16].

*Characterization of $Gd_2L7$ structure, dynamics and relaxation properties.* In Fig. 1 the crystal structure of $Gd_2L7$ is shown (unpublished data) and compared with the structure predicted for GdDTPA-BMA. For analyzing the paramagnetic perturbations induced on $^1H$-$^{13}C$ HSQC signals of α-BTX by $Gd_2L7$, a larger and more hydrophobic probe than GdDTPA-BMA, measurements of relaxivity, $r_{1p}$, at 298 K and pH=7.2 have been carried out for the two gadolinium probes. $r_{1p}$ values of 2.9 and 7.2 $mM^{-1} s^{-1}$ were obtained at 600 MHz for GdDTPA-BMA and $Gd_2L7$, respectively. The fact that $Gd_2L7$ relaxivity is more than twice higher than that of GdDTPA-BMA is due not only to the presence of two paramagnetic centers but also to its larger size that results in a significantly slower rotational dynamics in solution. The water solutions of $Gd_2L7$ and GdDTPA-BMA have been also investigated by $^1H$ and $^{17}O$ NMR relaxometric techniques in order to understand the details of their interactions with solvent molecules. Measurements of the temperature dependence of the water $^{17}O$ transverse relaxation rate, $R_{2p}$, represent a well-established procedure for an accurate determination of the coordinated water exchange dynamics. In fact, the dominant mechanism is the modulation of the Gd-$^{17}O_w$ scalar interaction which depends directly on the rate of water exchange, $k_{ex}$ ($k_{ex}=1/\tau_M$), and on the electronic relaxation times $T_{1,2e}$ [17]. The $R_{2p}$ data are analyzed with the use of the Swift-Connick equations that depends on several parameters including: i) the parameters related to the electronic relaxation times $T_{1,2e}$ (i.e. the trace of the square of the zero-field splitting, $\Delta^2$); the correlation time describing the modulation of the zero-field splitting, $\tau_V$; and its activation energy, $E_v$; ii) the enthalpy, $\Delta H_M^{\#}$ of activation for the water exchange process; iii) the hyperfine Gd-$^{17}O_w$ coupling constant, $A/\hbar$. The experimental data for $Gd_2L7$ were measured at 2.1 T on a 0.065 M aqueous solution. The $R_{2p}$ values increase by increasing temperature up to ca. 330 K where they level off, with a behaviour rather similar to that measured for GdDTPA-BMA and typical of systems in slow exchange (long value of $\tau_M$). The parameters obtained from the fit of the data to the standard theory are reported in Table 1.

**Table 1**
Selected best-fit parameters obtained from the analysis of the $^{17}$O NMR data (2.1 T) of Gd$_2$L7 (65 mM; pH = 6.2)

| Parameter | **Gd$_2$·L7** | Gd-DTPA-BMA[a] |
|---|---|---|
| $\Delta^2$ (s$^{-2}$·10$^{-19}$) | 5.0 ± 0.3 | 4.1 ± 0.2 |
| $^{298}\tau_V$ (ps) | 19 ± 1 | 25 ± 1 |
| $E_V$ (kJ mol$^{-1}$) | 6.0 ± 0.9 | 3.9 ± 1.4 |
| $^{298}\tau_M$ (μs) | 3.0 ± 0.2 | 2.2 ± 0.1 |
| $\Delta H^{\#}_M$ (kJ mol$^{-1}$) | 48.0 ± 0.4 | 47.6 ± 1.1 |
| $A/\hbar$ (10$^6$ rad s$^{-1}$) | -4.1 ± 0.3 | -3.8 ± 0.2 |
| $q^b$ | 1 | 1 |

[a] From reference 21; [b] Fixed during the least-squares procedure

The mean residence lifetime $\tau_M$ at 298 K for Gd$_2$L7 (3.0 μs) is sensibly longer than that measured for GdDTPA-BMA (2.2 μs) [18] and one of the longest reported so far for a neutral Gd-complex.

The knowledge of $\tau_M$ allows the evaluation of the other relaxation parameters by the measurement and analysis of the magnetic field dependence of the proton relaxivity on a fast-field cycling relaxometer over an extended frequency range (0.01-70 MHz). This was done for Gd$_2$L7 at 298 K and pH = 6.2. The relaxivity is constant from 0.01 to ca. 2 MHz, showing a dispersion around 6 MHz and remaining nearly constant at high fields (10-100 MHz). The shape and amplitude of the curve obtained is consistent with a dimeric species where each Gd center is coordinated by a single water molecule ($q$=1) and it is also clear that the relaxivity at low fields is partially limited by the long $\tau_M$ value, as for GdDTPA-BMA. The best-fit parameters, obtained from the analysis of the data in terms of the Solomon-Bloembergen-Morgan equations for the inner-sphere relaxation mechanism and Freed's equation for the outer sphere component [19]. The relevant results are the sensibly longer rotational correlation time, $\tau_R$, and distance of closest approach of the outer sphere water molecules, $a$, as compared to GdDTPA-BMA, whereas their electronic relaxation times are rather similar.

Since the obtained $r_{1,p}$ values are strictly dependent on the exchange rate of the water molecule coordinated to the Gd(III) ion, the optimal Gd$_2$L7 concentration to achieve α-BTX NMR signal attenuations similar to the ones previously obtained in presence of 2mM GdDTPA-BMA [14] was empirically evaluated by adding increasing amount of a Gd$_2$L7 solution to the protein sample. Thus, a final Gd$_2$L7 concentration of 0.5 mM was chosen for the present investigation.

*Paramagnetic attenuation analysis of α-BTX NMR signals.* In 0.5 mM and 2.0 mM paramagnetic solutions respectively of Gd$_2$L7 and GdDTPA-BMA, α-BTX protein signals appear similarly broadened even though the water signal is much wider in the second case. The different Gd$_2$L7 and GdDTPA-BMA relaxivities, as well as their interaction with water molecules, account for this finding. It should be noted here that this $Gd_2L7$ feature is in principle very favorable for quantitative volume determinations of peaks lying close to the solvent signal.

In Fig. 3 the fingerprint region of α-BTX $^1$H-$^{13}$C HSQC spectra, recorded in the absence and in the presence of $Gd_2L7$, is shown. Peak volumes of 45 well resolved CαH correlations could be quantified with enough accuracy to make speculations on signal attenuations, $A_{iGd2}$. A comparative analysis of the latter $A_i$'s with the ones previously measured for α-BTX CαH signals in the presence of GdDTPA-BMA [14], $A_{iGd}$, is shown in Fig. 2b. $A_{iGd2}$'s and $A_{iGd}$'s are, in general very similar, being their average difference $\Delta=0.15$ and with a $\sigma_\Delta=0.11$. To discuss the extent of paramagnetic attenuations, the standard deviation of $A_{iGd2}$'s, $\sigma_A$, has been calculated. Since a $\sigma_A$ value equal to 0.36 was found, strong paramagnetic effects have been assessed only for those CαH correlations whose $A_{iGd2}$ values were larger than 1.36, as in the case of Ala31, Ser34, Arg36, Gly37 and Val40. It is interesting to note that the improved quality of the $^1$H-$^{13}$C HSQC spectrum obtained in the presence of $Gd_2L7$, comparing the one obtained in the presence of GdDTPA-BMA, allows an accurate peak volume determination for six more correlations, including the one of Val40 CαH, very attenuated and not determined in our previous investigation [14].

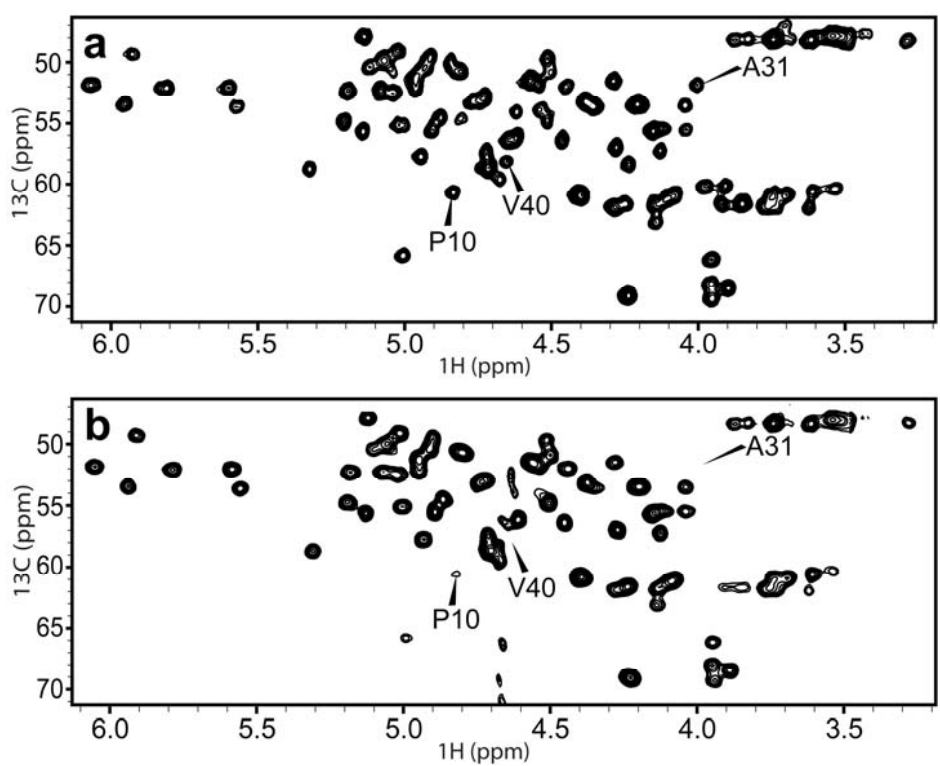

Figure 3: The fingerprint region of α-bungarotoxin $^1$H-$^{13}$C HSQC spectra recorded (a) in the absence and (b) in the presence of 0.5 mM $Gd_2L7$.

Only few CαH correlations exhibit $A_{iGd2}$ values below 0.64, *i.e.* smaller than the average $A_{iGd2}$ by a factor of σ. The small size of the neurotoxin and the long distance perturbation effect of the paramagnetic probe [20] account for this finding.

**Discussion**

The high resolution structure of the complex between α-BTX and the extracellular domain of the α1 subunit of AchR, α211, has been recently resolved [15] and it can be used as a reference for discussing the extent of $Gd_2L7$ induced perturbations to the α-BTX methyne groups. On the basis of the Protein Data Bank [21] structure the α-BTX-α211 complex, PBD code 2QC1, it is apparent that the highest $A_i$ values are all grouped in the toxin side where the binding with α211 occurs, see Fig. 4. Conversely, all the surface exposed α-BTX methynes which experience only limited paramagnetic attenuations are located in a wide region which is almost opposite to the protein binding site.

As already suggested [14], for a structural analysis of the obtained paramagnetic perturbations, atom depths, rather than accessible surface area, ASA, should be considered. Atom depth, indeed, reflects more properly than ASA the through-space character of those dipolar interactions between electronic and nuclear spins which generate the observed paramagnetic perturbations. In facts, backbone nuclei which are very close to the surface and, therefore, very approachable by paramagnetic probes, might also exhibit ASA=0, as in the frequent case where they are shielded by their neighboring side chain atoms.

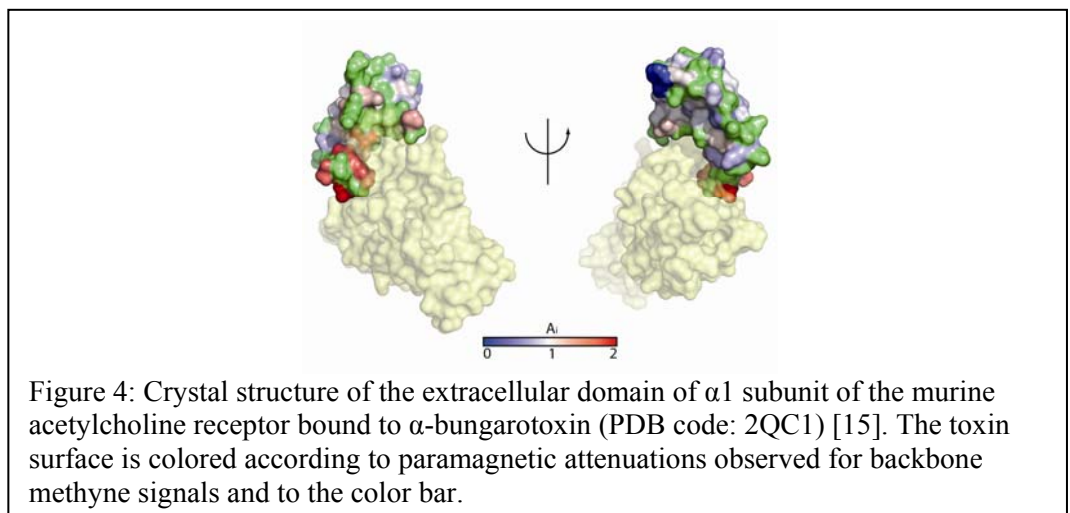

Figure 4: Crystal structure of the extracellular domain of α1 subunit of the murine acetylcholine receptor bound to α-bungarotoxin (PDB code: 2QC1) [15]. The toxin surface is colored according to paramagnetic attenuations observed for backbone methyne signals and to the color bar.

Atom depths can be efficiently quantified by calculating the surface exposed volume of spheres having a suitable radius and centered on the investigated atoms [22]. Thus, on the basis of the used reference structure, atom depth indexes, $D_i$, of each α-BTX atom can be calculated, see Fig. 2c and the higher is the $D_i$ value the smaller is the distance between a given atom and the accessible surface. The comparative analysis of $D_i$, rmsf and $A_i$ values, suggested in Fig. 2, points out that high flexibility and close surface proximity of α-BTX backbone are needed to observe strong paramagnetic attenuations. Data shown in Fig. 2 suggest also that other contributions are present in controlling the approaches of the used paramagnetic probes to the toxin surface, since other flexible and surface exposed protein moieties are characterized by average or low $A_i$ values. This is the case of the so called α-BTX finger I and III regions, respectively residues 3-12 and 47-57. The predicted presence of high density water could explain a more hindered approach of both paramagnets to the latter surface regions [14]. The accessibility of finger III could experience an additional hindrance due to the involvement of this protein moiety in the formation of dimeric species seen both in the crystal state [23] and in solution [24].

The strong similarity of $A_{iGd}$ and $A_{iGd2}$ values cannot be attributed to a mere coincidence of specific protein-probe interactions, but rather to the characteristics of α-BTX surface dynamics. Furthermore, it should be noted that α-BTX structural features identify two protein sides: side A, characterized by i) reduced backbone flexibility, ii) presence of many high density hydration sites, as predicted by MD simulations [14] and iii) protein surface scarcely approached by the two gadolinium complexes. A complementary trend of the above mentioned features can be observed in the α-BTX side B, the one which incorporates the protein active site.

It can be concluded that paramagnetic perturbations, induced by soluble and neutral Gd complexes on suitable NMR signals, may represent a powerful source of information to find general rules which may explain what makes a specific protein region as the preferentially accessible one. The data discussed here suggest also that the ditopic probe, causing smaller water signal broadening than GdDTPA-BMA, due to its longer $\tau_M$, seems to be very suitable for protein accessibility investigations. The induced paramagnetic perturbations can be, indeed, analyzed in spectra where better water suppression can be achieved and more information can be derived. For a further reduction of water signal broadening, new Gd(III) based paramagnetic probes, where $H_2O$ cannot exchange from the inner ion coordination sphere to the bulk solvent ($q=0$ systems), are now under consideration. A large variety of new relaxation reagents matching the requirements for probing molecular surface accessibility should be available to find unambiguous correlations between paramagnetic perturbation profiles with protein hydration and flexibility.

**Experimental Section**

*Sample preparation and NMR measurements*

α-BTX were obtained from Sigma and used without any further manipulation. $Gd_2L7$ was prepared as previously described [3]. NMR measurements, run at 303 K and pH 6.0 to reproduce the experimental conditions of the original structural study of α-BTX, were obtained with a Bruker Avance 600 spectrometer. Data processing was performed with the NMRPipe software [25]. The $^1H$ chemical shifts were referenced on trimethylsilylpropionic 2,2,3,3-d$_4$ acid sodium salt (TSP) at 0 ppm. NMR samples were prepared by dissolving the protein in H$_2$O/D$_2$O (95:5) to obtain a 1 mM solution of the protein. The paramagnetic NMR samples contained 0.5 mM Gd$_2$L7, an optimal probe concentration to observe sizeable signal attenuations.

To compare cross peak volumes, $V_i$, obtained in different set of experiments and measured with a confidence higher than 90%, their autoscaled values $\upsilon_i$ have been used according to the relation [26]:

$$\upsilon_i^{p,d} = \frac{V_i^{p,d}}{(1/n)\sum_n V_i^{p,d}}$$

Paramagnetic attenuations, A$_i$, were calculated from the autoscaled diamagnetic and paramagnetic peak volumes, respectively $\upsilon^d$ and $\upsilon^p$, according to the relation [26]:

$$A_i = 2 - \frac{\upsilon_i^p}{\upsilon_i^d}$$

For each GLY residue both the Cα-Hα1 and Cα-Hα2 correlations were analyzed and the averaged A$_i$ is reported.

Data analysis was performed with Sparky software (http://www.cgl.ucsf.edu/home/sparky).

*Relaxivity measurements*

The proton 1/$T_1$ NMRD profiles were measured on a Stelar fast field-cycling FFC-2000 (Mede, Pv, Italy) relaxometer on about 0.25-1.0 mmol gadolinium solutions in non-deuterated water. The relaxometer operates under computer control with an absolute uncertainty in 1/$T_1$ of ± 1%. The NMRD profiles were measured in the range of magnetic fields from 0.00024 to 1.6 T (corresponding to 0.01–70 MHz proton Larmor frequencies). Additional experimental data (300 and 600 MHz) were acquired on high resolution Bruker Avance NMR spectrometers and compared with those calculated from the fitting of the NMRD profiles.

Variable-temperature $^{17}$O NMR measurements were recorded on a JEOL EX-90 (2.1 T) spectrometer equipped with a 5 mm probe. Solutions containing 2.0% of the $^{17}$O isotope (Cambridge Isotope) were used. The observed transverse relaxation rates were calculated from the signal width at half-height. Other details of the instrumentation, experimental methods, and data analysis have been previously reported [27].

*Depth index calculation*

The 3D atom depths, reported as depth indexes $D_{i,r}$'s, were calculated on the basis of the α-BTX solution structure (PDB code: 1IK8) [28].

$D_{i,r}$ is defined according to the relation [22]:

$$D_{i,r} = \frac{2V_{i,r}}{V_{0,r}},$$

where $V_{i,r}$ is the exposed volume of a sphere of radius r (sampling radius) centered on atom i and $V_{0,r}$ is the total volume of the same sphere. The sampling radius was set to 8 Å. The probe radius, used for the computation of the macromolecular surface, was set to 1.4 Å corresponding to the water molecular radius. For each C-H correlation, the averaged $D_{i,8}$ is reported.

*Molecular Dynamics simulation*

A 16 ns MD simulation was performed starting from the α-BTX NMR structure (PDB code: 1IK8), using the GROMOS force field [29] and the GROMACS package [30]. Details on the MD run are supplied elsewhere [14]. All the analyses of MD trajectory were carried out using standard GROMACS tools.

*Structure prediction of* GdDTPA-BMA. On the basis of the crystal structure of GdDTPA-BEA [31], the structure of the homologous complex GdDTPA-BMA has been modeled by using MOPAC [32] and reported in Fig. 1.

**Acknowledgements**

Thanks are due to the University of Siena and University of Eastern Piedmont for financial support. W.-T.W. thanks the Hong Kong Research Grants Council (HKU7116/02P) and the University of Hong Kong for financial support. A.P.-L.T. thanks The University of Hong Kong for financial support.